\documentclass{aa}  

\usepackage{graphicx}
\usepackage{txfonts}
\newcommand{\chandra}{{\it Chandra}}
\newcommand{\fermi}{{\it Fermi}}

\begin{document}

   \title{A high-resolution radio morphology and polarization of the kpc-scale X-ray jet of PKS\,1127$-$145}

   \author{M. Orienti
          \inst{1}\
          \and
          A. Siemiginowska\inst{2}
          \and
          F. D'Ammando\inst{1}
          \and
          G. Migliori\inst{1}
}
   \institute{Istituto di Radioastronomia - INAF, Via P. Gobetti 101, I-40129 Bologna, Italy\\
              \email{orienti@ira.inaf.it}
         \and
             Harvard Smithsonian Center for Astrophysics, 60 Garden St, Cambridge, MA 02138, USA
             }

   \date{Received <date> / Accepted <date>}

% \abstract{}{}{}{}{}
% 5 {} token are mandatory
 
  \abstract
   {We report on new multi-frequency Very Large Array (VLA) radio observations and {\it Chandra}
  X-ray observations of a radio-loud quasar with a $\sim$300\,kpc long jet, PKS\,1127$-$145, during a flaring event detected in $\gamma$-rays by the {\it Fermi} Large Area Telescope in 2020 December. The high angular resolution of the new radio images allows us to disentangle for the first time the inner kpc-scale jet from the core contribution. The inner radio jet, up to 15 kpc from the core, is highly polarized (33 per cent) and the magnetic field is parallel to the jet axis. At about 18 arcsec from the core the jet slightly bends and we observe a re-brightening of the radio emission and a 90-degree rotation of the magnetic field, likely highlighting the presence of a shock that is compressing the magnetic field to a plane perpendicular to the jet axis and where efficient particle acceleration takes place. At the same position the X-ray emission fades, suggesting a deceleration of the bulk velocity of the jet after the bend. A change in velocity and collimation of the jet is supported by the widening of the jet profile and the detection of a limb-brightened structure connecting the bending region with the jet termination. The limb-brightened structure might indicate the co-existence of both longitudinal and transverse velocity gradients at the jet bending.
  There is no evidence for significant brightening of the kpc-scale
  jet in the radio or X-ray band during the $\gamma$-ray flare. The
  X-ray flux, $F_{\rm 2-10\,keV} =
  (6.24\pm0.57)\times10^{-12}$\,ergs~s$^{-1}$~cm$^{-2}$, measured by
  {\it Chandra} from the quasar core is consistent with the flux measured by the X-Ray Telescope on board the {\it Neil Gehrels Swift Observatory} after the high-energy flare. Our results indicate that the $\gamma$-ray flaring region is located within the VLA source core.  }

\titlerunning{Kpc-scale X-ray jet of PKS\,1127-145}
\authorrunning{M. Orienti et al.}
   \maketitle

%
%________________________________________________________________

\section{Introduction}

  \label{sec:intro}

The extragalactic $\gamma$-ray sky is dominated by blazars. The emission of this class of active galactic nuclei (AGN) comes mainly from the relativistic jet that is aligned close to our line of sight. As a consequence the luminosity is augmented by Doppler boosting and beaming effects and variability is observed across the electromagnetic spectrum. Among the extragalactic $\gamma$-ray sources detected by the Large Area Telescope (LAT) on board the {\it Fermi Gamma-ray Space Telescope} satellite (hereafter \fermi), flat spectrum radio quasars (FSRQ) show the most dramatic flaring events. Their $\gamma$-ray flux may increase by more than an order of magnitude than the average level, with a doubling-time of a few hours or even shorter \citep[e.g.,][]{abdo11,hayashida15,dammando19}.

Despite decades of studies, many aspects of the high energy emission from AGN are still elusive. Among them, the location of the high-energy emitting region and the main radiative processes at work have been investigated intensively by multi-band (and recently by multi-messenger) observations. 
The detection of a (sub-)hour variability by {\it Fermi}-LAT of some FSRQs suggests a location between the broad line region (BLR) and the molecular torus \citep[e.g.,][]{abdo11,hayashida15,ackermann16,acharyya21}. On the other hand, 
\citet{costamante18} could not find significant evidence of cut-off signatures at high energies compatible with $\gamma$-$\gamma$ interactions with BLR photons in the $\gamma$-ray spectra of a sample of FSRQs, suggesting that the $\gamma$-ray emitting region is far beyond the BLR. Very long baseline interferometry (VLBI) observations point out the appearance of superluminal jet components close in time with some $\gamma$-ray flares, indicating the radio core as the locus of high-energy emission, i.e. a few pc from the central engine \citep[e.g.,][]{marscher10,agudo11,mo13,jorstad17}.

Not all the high-energy flares originate in the same region even when the same source is considered \citep[see, e.g.,][]{marscher08,marscher10,mo13}. A remarkable example is the radio galaxy M\,87. The high resolution of radio and X-rays observations discovered the jet knot HST-1, at 120 pc from the AGN, as the locus of the high-energy emission observed in 2005 \citep{cheung07,harris2009}. Instead, the high activity state observed in 2012 originated at the source core, whereas HST-1 remained quiescent \citep{hada14}. 
Another example is the X-ray flaring activity observed in the extended jet of Pictor\,A that originated at about 48 arcsec ($\sim$33 kpc) from the core, indicating that variability can be observed in the outer regions of relativistic jets \citep{marshall10,hardcastle16a}. 

X-ray variability in kpc-scale jets and hotspots on timescales of a few months to years does not seem to be so uncommon \citep{meyer23}. Such relatively short time scales challenge a simple model of inverse Compton (IC) scattering of the cosmic microwave background (CMB) photons, and favour synchrotron emission from a second highly-energetic population of relativistic electrons in compact (pc-scale size) regions \citep[e.g.,][]{hardcastle04,tingay08,hardcastle16a,migliori20,meyer23}. The IC-CMB model is called into question by other observational evidence, like the detection of kpc-scale displacement between X-ray and radio emission in several knots and hotspots \citep[e.g.,][]{hardcastle07,mingo17,mo20,migliori20,reddy23}. However, it should be kept in mind that high-redshift jets may be different from those at low and intermediate redshift \citep[e.g.,][]{mckeough16,ighina22}.

Not many AGN have relativistic jets that can be imaged on arcsecond scale in X-rays. Moreover, only a handful of them are $\gamma$-ray emitters. Multi-wavelength studies of these jets are crucial for investigating particle acceleration and emission mechanisms far away from the central engine. 

PKS\,1127$-$145, at $z$ = 1.187 \citep{drinkwater97}, is one of the few $\gamma$-ray-emitting FSRQs with a prominent X-ray jet extending for $\sim$30\arcsec \citep{siemiginowska02}.  
This is one of the longest jets observed so far in X-rays. It was discovered in the first observation of PKS\,1127$-$145 with the {\it Chandra} X-ray Observatory (hereafter \chandra). 
Three main knots are detected in X-rays along the inner part of the jet, whereas the radio emission peaks at the two outer knots, associated with weak X-ray emission. The misalignment between radio and X-ray emission challenges our understanding of the jet structure and dominant radiation mechanism at the origin of the high-energy emission \citep{siemiginowska02,siemiginowska07}.

\fermi-LAT observed enhanced $\gamma$-ray activity from PKS\,1127$-$145 on 2020 December reaching a daily $\gamma$-ray flux (E $>$ 100 MeV) of (1.6 $\pm$ 0.3)$\times$ 10$^{-6}$ photons cm$^{-2}$ s$^{-1}$ on December 10 \citep{angioni20}, corresponding to a flux increase of a factor of about 50 relative to the value reported in the fourth {\em Fermi}-LAT source catalogue \citep{abdollahi20}. That was the first strong $\gamma$-ray flaring activity observed from this source in the {\it Fermi} era. Follow-up {\it Swift}-XRT observations on December 13 found that the X-ray flux of the source increased as well, reaching the highest X-ray
flux observed by {\it Swift}-XRT so far \citep{dammando20a}, suggesting a physical connection between the $\gamma$-ray and X-ray radiation processes. The subsequent {\it Swift}-XRT observations performed on December 15 and 17 confirmed that the X-ray flare was continuing \citep{dammando20b}. However, the resolution of {\it Swift}-XRT is not adequate to spatially resolve the X-ray emission and locate the flaring region. 

In this paper we present results on new {\it Chandra} X-rays and Jansky Very Large Array (VLA) radio observations of PKS\,1127$-$145 performed just after the $\gamma$-ray flaring event. These observations aim at investigating the jet structure and the location of the X-ray flaring region. Full-polarization radio data obtained with the VLA in A-configuration enable us to study for the first time the magnetic field structure with high (sub-arcsecond) angular resolution along the kpc-scale jet of PKS\,1127$-$145.

This paper is organized as follows: Section~\ref{sec:obs} describes the setting of the observations; results on the kpc-scale jet morphology in radio and X-rays are presented in Section~\ref{sec:results} and discussed in Section~\ref{sec:discuss}. We draw our conclusions in Section~\ref{sec:con}.

Throughout this paper, we assume the following cosmology: $H_{0} =
70\; {\rm km\,s^{-1}\, Mpc^{-1}}$, $\Omega_{\rm M} = 0.27$ and $\Omega_{\rm \Lambda} = 0.73$, in a flat Universe. At the redshift of the source, $z$ = 1.187, 1\,arcsec
corresponds to 8.4\,kpc \citep{wright06}. The spectral index is defined as 
$S {\rm (\nu)} \propto \nu^{- \alpha}$. The position angle is measured from
North to East, where North is up and East is left.

\section{Observations}
\label{sec:obs}

\subsection{VLA observations and data analysis}
\label{sec:vla}

We were awarded 4.5 hr of Director's Discretionary Time (DDT) at the VLA (project
code VLA/20B-460) to observe
PKS\,1127$-$145 after the detection of a flaring state in $\gamma$ rays
\citep{angioni20}. VLA observations were performed on
2021 January 7 in L (1-2 GHz), C (4-8 GHz), and
X (8-12 GHz) bands in full polarization mode when the array was in A configuration. On-source observing time was about 45 min in L and C bands, and 70 min in X bands, spread into several scans and cycling through frequencies in order to improve the {\it uv}-coverage.

\indent The source 3C\,286 was used as primary calibrator, band pass
calibrator, and electric vector position angle (EVPA) calibrator,
while the unpolarized source OQ\,208 was observed as D-term
calibrator. 

Calibration was performed using Common Astronomical
Software Applications (CASA) version 5.4.1
\citep{mcmullin17} following the standard procedure for VLA
data. Data were inspected and hanning smoothed to reduce Gibbs ringing produced by strong radio frequency interference (RFI) present in some spectral windows, mainly in L and C bands. After an initial flagging on bad data we calibrated the data sets. We checked for antenna position corrections and ionospheric total electron content corrections. We set the flux density model for 3C\,286 (no polarization model is set at this stage) using the \citet{perley17} scale. We then performed an initial delay and bandpass calibration using 3C\,286 before running a second flagging of RFI (setting '{\tt RFLAG}' in the CASA task {\tt FLAGDATA}). An initial gain calibration (both phase and amplitude) is performed. After applying the initial calibration to the data we did a further RFI flagging. Then we performed all the calibration again on the flagged data (delay, bandpass, gain calibration). 

At this point we performed the polarization calibration. First, we set the polarization model for 3C\,286\footnote{https://science.nrao.edu/facilities/vla/docs/manuals/obsguide/modes/pol.}. Then, we solved for the cross-hand delays for 3C\,286, before determining the D-terms for the unpolarized and unresolved calibrator OQ\,208. Last, the polarization angle was calibrated making use of 3C\,286.

Errors on the amplitude calibration, $\sigma_{\rm cal}$, were estimated by checking
the scatter of amplitude gain factors, and turned out to be about 3 per
cent in all bands, in agreement with the errors reported in
\citet{perley17}. Errors on the polarization angle are about 3-5 deg.

After the a-priori calibration we produced images using the CASA task {\tt tclean} with multi-termi multi-frequency synthesis deconvolution (nterms=2), Briggs weightings and robust=0.5. Before creating the final images we performed a few
phase-only self-calibration iterations decreasing the solution intervals from 60 seconds to 10 seconds, followed by a single amplitude self-calibration with a solution interval of the scan length \citep[see e.g.,][]{cornwell99}.

In addition to the total intensity images, we produced polarization
intensity and polarization angle maps
combining images in Stokes Q and U using the CASA task {\tt immath}. Pixels with values below 3 times the rms measured on the input images were masked.\\
\indent The restoring beam of the final images is 1.78$\times$1.11
arcsec$^2$ with a major axis position angle (PA) 17$^{\circ}$ at 1.5 GHz, 0.47$\times$0.29 arcsec$^{2}$ with PA 22$^{\circ}$ at 6 GHz, and 0.28$\times$0.17 arcsec$^{2}$ with PA 20$^{\circ}$ at 10 GHz.

We measure the flux density of the unresolved components
using the CASA task {\tt imfit} which performs a
two-dimensional Gaussian fit on the image plane. For resolved
components and for estimating the total flux density we use the task
{\tt viewer} which extracts the flux density on a
selected polygonal area on the image plane. Flux densities are
reported in Table \ref{vla_flux}. The polarized flux density is measured
on the same region as the one considered for total intensity
measurements and is reported in Table \ref{polla}, together with the EVPA. 

Errors on the total intensity and polarized flux densities are estimated by $\sigma = \sqrt{\sigma_{\rm cal}^2 + \sigma_{\rm rms}^2}$, where $\sigma_{\rm cal}$ is the error on the amplitude calibration, and $\sigma_{\rm rms}$ is the 1-$\sigma$ noise level of the rms measured on the image plane. The latter contribution depends on the area of the selected region used for extracting the flux density, $\theta_{\rm source}$, as 
$\sigma_{\rm rms} = {\rm rms} \times \sqrt{ \theta_{\rm source}/ \theta_{\rm beam}}$, where $\theta_{\rm beam}$ is the area of the Gaussian restoring beam.

Depending on the position on the image of the component considered, the rms is measured either far or close to the core component. The off-source noise level
of the final images is 0.1 mJy beam$^{-1}$, 0.025 mJy beam$^{-1}$, and
0.015 mJy beam$^{-1}$ in L, C and X bands, respectively. Imaging artifacts are
present close to the core component, and the rms is higher in that
area. This is likely due to the combination of the bright
core, the low declination of the source, and the relatively short
observing time. To improve the signal-to-noise ratio (S/N) we created a data set in which we subtracted the model visibility data of the bright core from the corrected visibility data with the CASA task {\tt uvsub}, leaving only the residuals. However, not all the artifacts could be removed and the rms did not improve significantly.

\begin{figure}
\begin{center}
\includegraphics[width=0.85\columnwidth]{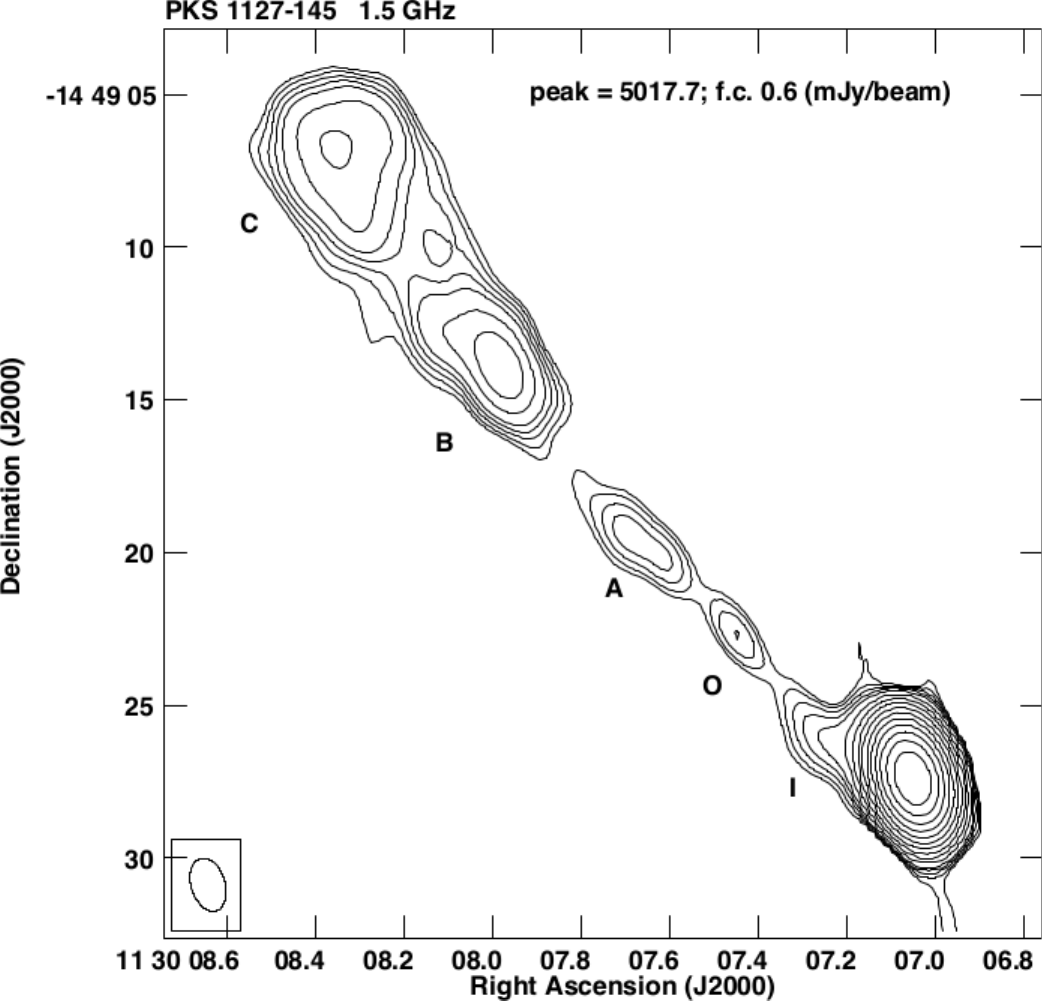}
\includegraphics[width=0.85\columnwidth]{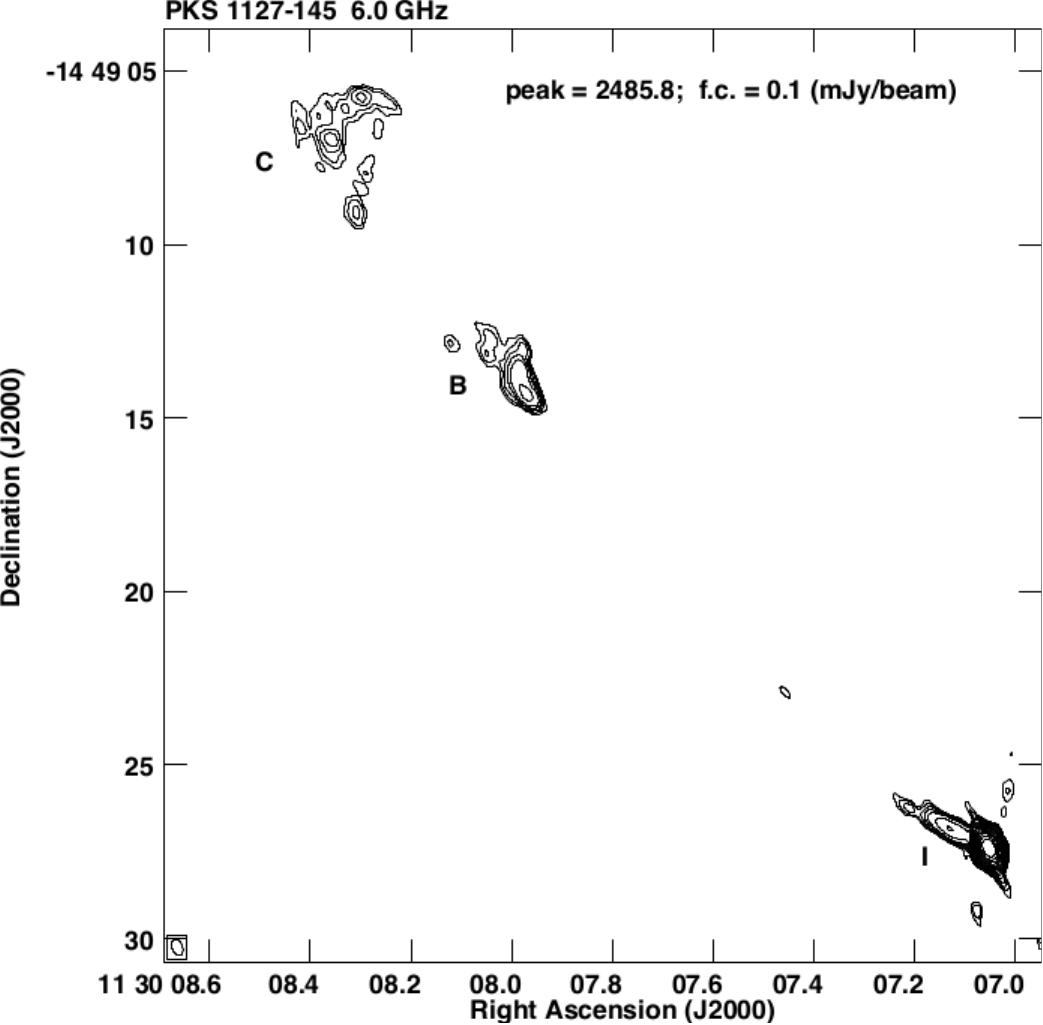}
\includegraphics[width=0.85\columnwidth]{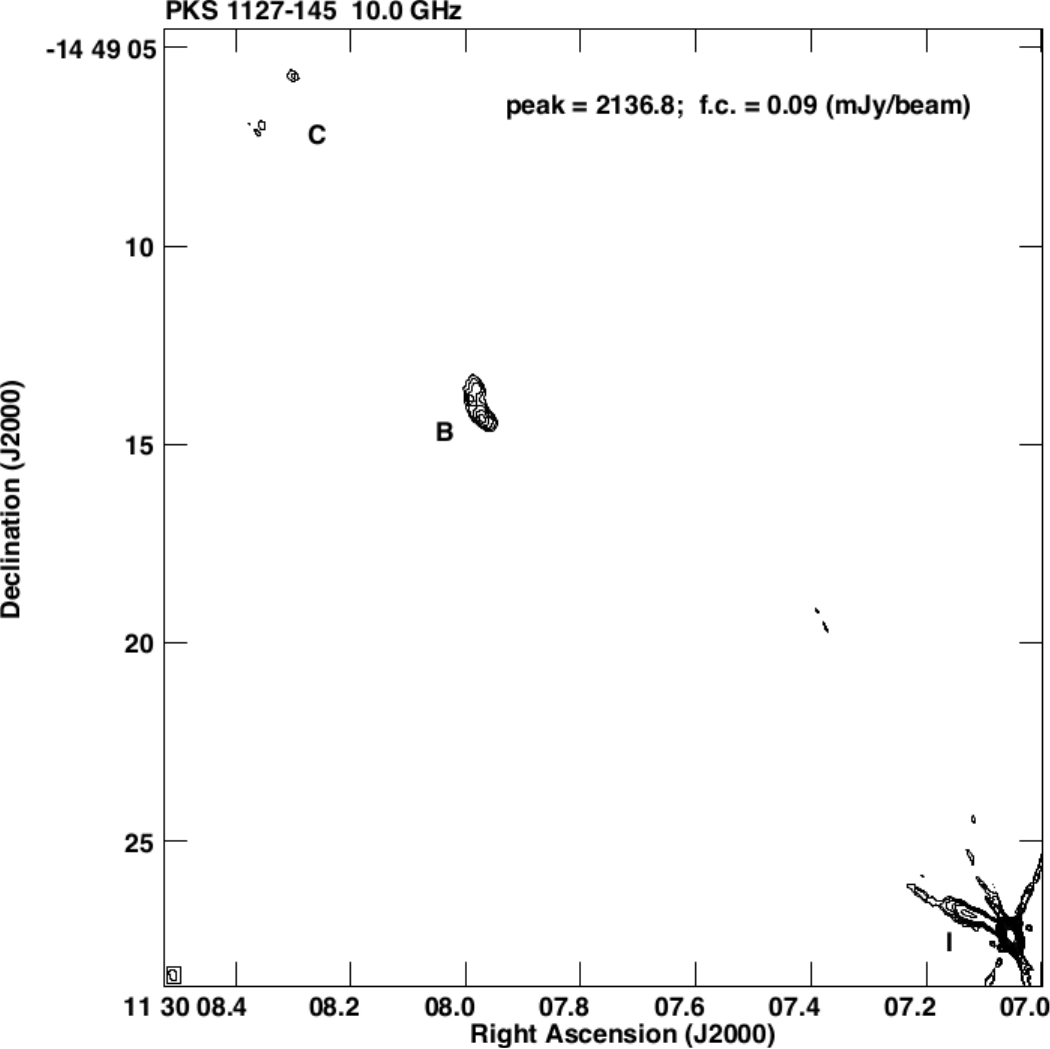}
\caption{Total intensity images of PKS\,1127$-$145 at 1.5 GHz ({\it top}), at 6 GHz ({\it middle}), and at 10 GHz ({\it bottom}). The first contour is 0.6, 0.1, and 0.075 mJy beam$^{-1}$ at 1.5, 6, and 10 GHz, respectively, and corresponds to three times the rms measured on the image plane close to the centre. Contours are drawn at [-1, 1, 1.4, 2, 2.8, 4, 5.6, 8, 16, 32, 64, ...] times the first contour. The restoring beam is plotted in the bottom left-hand corner of each image.}
\label{vla-image}
\end{center}
\end{figure}

Fits files of the final images were imported into the Astronomical Image Processing System (AIPS), where contour images were produced with the {\tt KNTR} task. Final images are shown in Fig. \ref{vla-image}.

\subsection{{\it Chandra} observations and data analysis}
\label{sec:chandra}

The \chandra\ DDT observation of PKS\,1127$-$145 was performed on 2021 January 1 during the time of flaring events detected by \fermi\ and {\it Swift}. Our goal was to
identify the X-ray site of the flare and check if the activity could
be located outside the quasar core. The \chandra\ point spread
function (PSF) allows for the best angular resolution X-ray images available today. Here we present this new observation together with the archival data in order to inspect for any variability of the jet. 

During the course of the mission \chandra\ observed PKS\,1127$-$145 three times (see Table~\ref{tab:chandra}) using the ACIS-S3 detector with the target located at the aimpoint on the back illuminated charge coupled device (CCD). The quasar is bright and in order to limit the CCD pileup all the observations were taken with 1/8 subarray readout. The data mode was set to VFAINT\footnote{For information on the {\it Chandra} data see: https://cxc.harvard.edu/proposer/POG/html/index.html} in the first two observations which typically improves identification of background events. The FAINT mode was used in the most recent observation.
The first two observations were obtained early in the mission with a good detector response across all the energies, from 0.3--8 keV. However, the most recent observation performed in 2021 January had a degraded sensitivity due to the ACIS-S contamination build up which significantly reduced the number of counts in the soft energies, i.e. below 1 keV.

We used CIAO version 4.15 software \citep{fruscione06} for data analysis and \texttt{Sherpa} for fitting and modeling \citep{freeman01,refsdal11}. We reprocessed all three observations using {\tt chandra\_repro} tool and applied the recent calibration products available in the CALDB\,v.4.10.4.

\begin{figure}
\begin{center}
\includegraphics[width=\columnwidth]{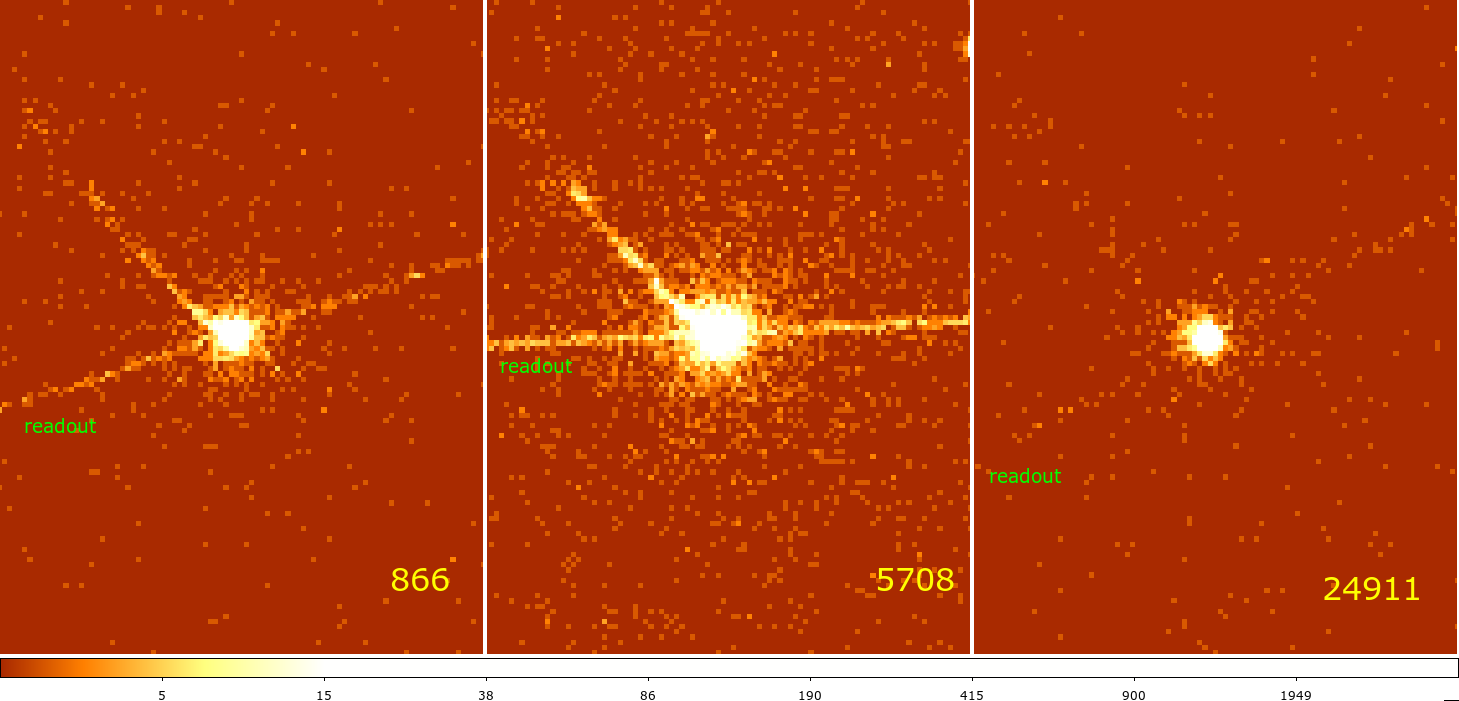}
\caption{ACIS-S 0.5-7\,keV image showing a total number of counts per pixel obtained in each \chandra\, observation. The \chandra\, ObsID number is indicated in the right bottom corner in each panel, from left to right: 866, 5708, 24911. The image pixels are native ACIS-S pixel size of 0.492\arcsec. The image is displayed in logarithmic scale with the color bar scale indicating a number of counts per pixel. Note that in addition to the jet the readout streak (labeled green) is visible with different angles dependent on the \chandra\, roll angle during the observation. }
    \label{fig:xray-obsid}
\end{center}
\end{figure}

\begin{figure}
\begin{center}
\includegraphics[width=\columnwidth]{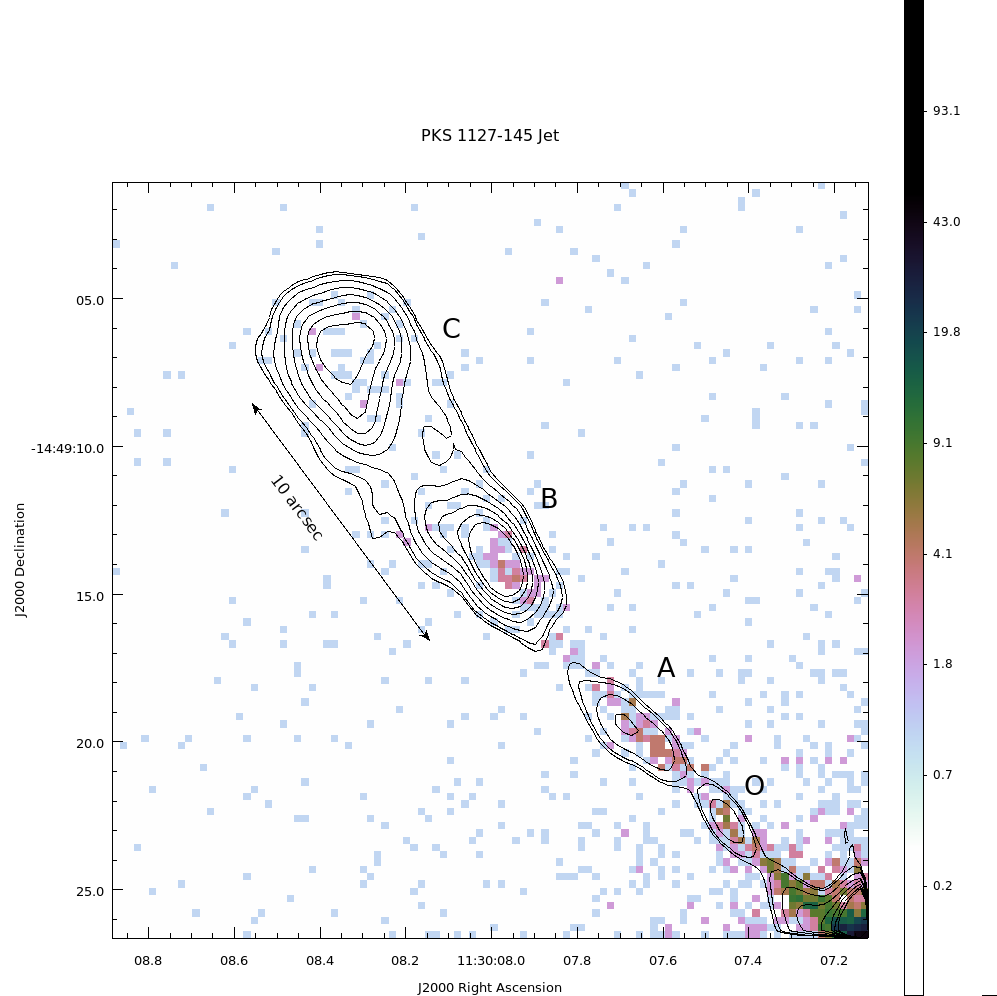}
\caption{ACIS-S 0.5-7\,keV mosaic image binned into 0.246\arcsec\, pixels (0.5 bin scale). The color bar shows the number of counts per pixel. The black contours are from 1.6\,GHz L-band VLA map starting at 0.45\,mJy beam$^{-1}$ and increasing by $\sqrt{2}$.
The main knots are labeled and the arrow marks 10\arcsec\, scale along the outer part of the jet. The quasar core is located at the low right corner just outside the main frame.}
    \label{fig:xray-jet}
\end{center}
\end{figure}

The standard aspect reconstruction shows a relatively large offset, of about 1\arcsec, between the first two (ObsID 866, 5708) and the most recent observation (ObsID 24911). We adjusted 866 and 24911 observations to match the coordinates of the longest 5708 observation and merged the three observations to obtain the best available \chandra\ image of the source. The images from individual observations are shown in Fig.~\ref{fig:xray-obsid} and the final merged image 
is shown in Fig.~\ref{fig:xray-jet}. The quasar is bright and the ACIS-S readout streaks are visible in the image presented in Fig.~\ref{fig:xray-obsid}, as we did not remove them at this stage. We note that the streak is much fainter in the most recent observation with the shorter exposure and degraded soft energy response. 

The jet of PKS\,1127$-$145 is quite prominent and the outermost structure can be analyzed in relation to the radio emission, including polarization present in the outer knots. In addition we perform the analysis of the innermost structures in the vicinity of a jet bend.

\begin{figure*}
\begin{center}
\includegraphics[width=0.95\columnwidth]{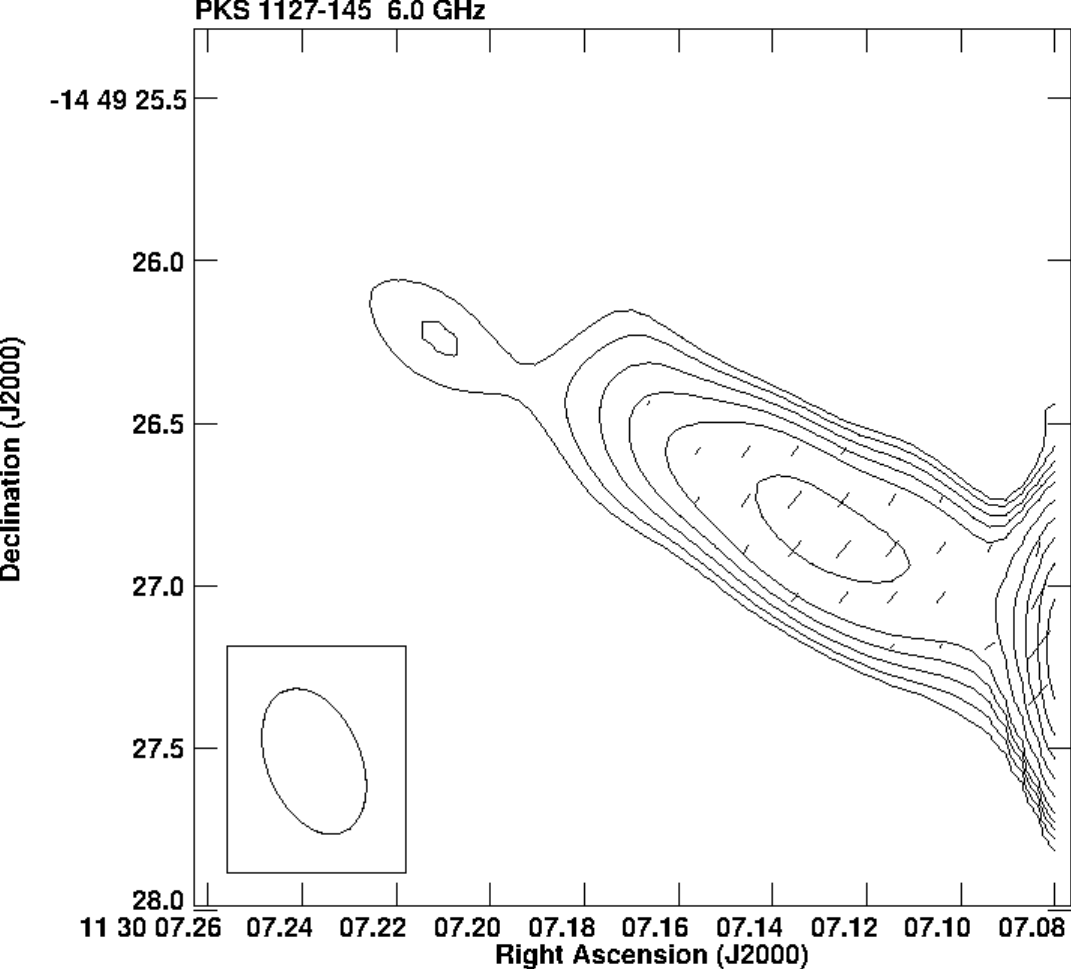}
\includegraphics[width=0.85\columnwidth]{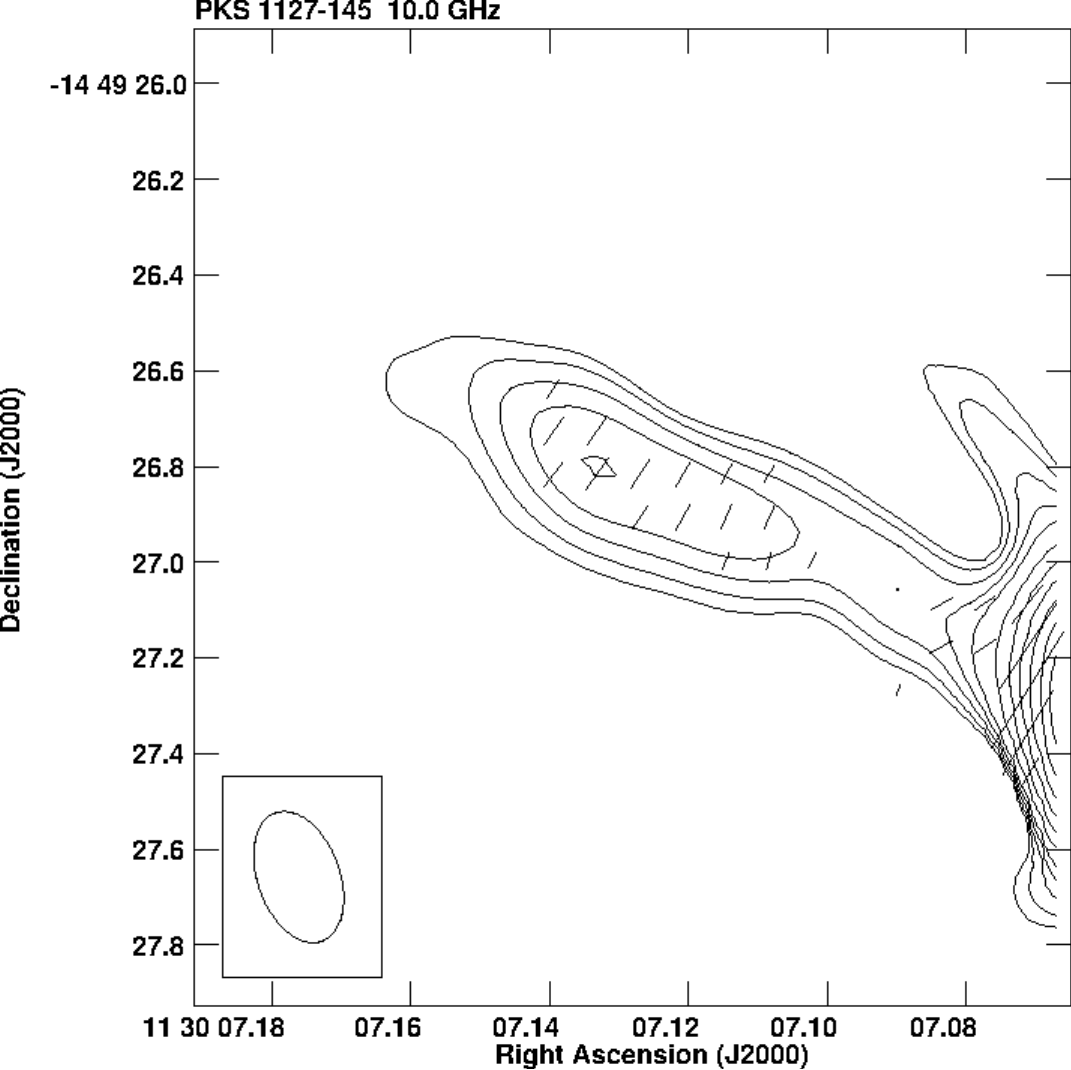}
\caption{VLA total intensity image of the inner 2-arcsecond jet at 6 GHz ({\it left})
  and at 10 GHz ({\it right}).
The first contour is 0.2 mJy beam$^{-1}$ at both frequencies. Contours are drawn at [-1,1,1.4,2,2.8,4,5.6,8,16,32,...] times the first contour. The restoring beam is plotted in the bottom left-hand corner. Vectors superimposed on the total intensity contours show the position angle of the electric vector, where 0.25-arcsec length corresponds to 2.5 and 0.85 mJy beam$^{-1}$ (polarization intensity) at 6 and 10 GHz, respectively. The quasar
core is located at the low right corner just outside the main frame.}
    \label{inner-image}
\end{center}
\end{figure*}

\begin{figure*}
\begin{center}
\includegraphics[width=\columnwidth]{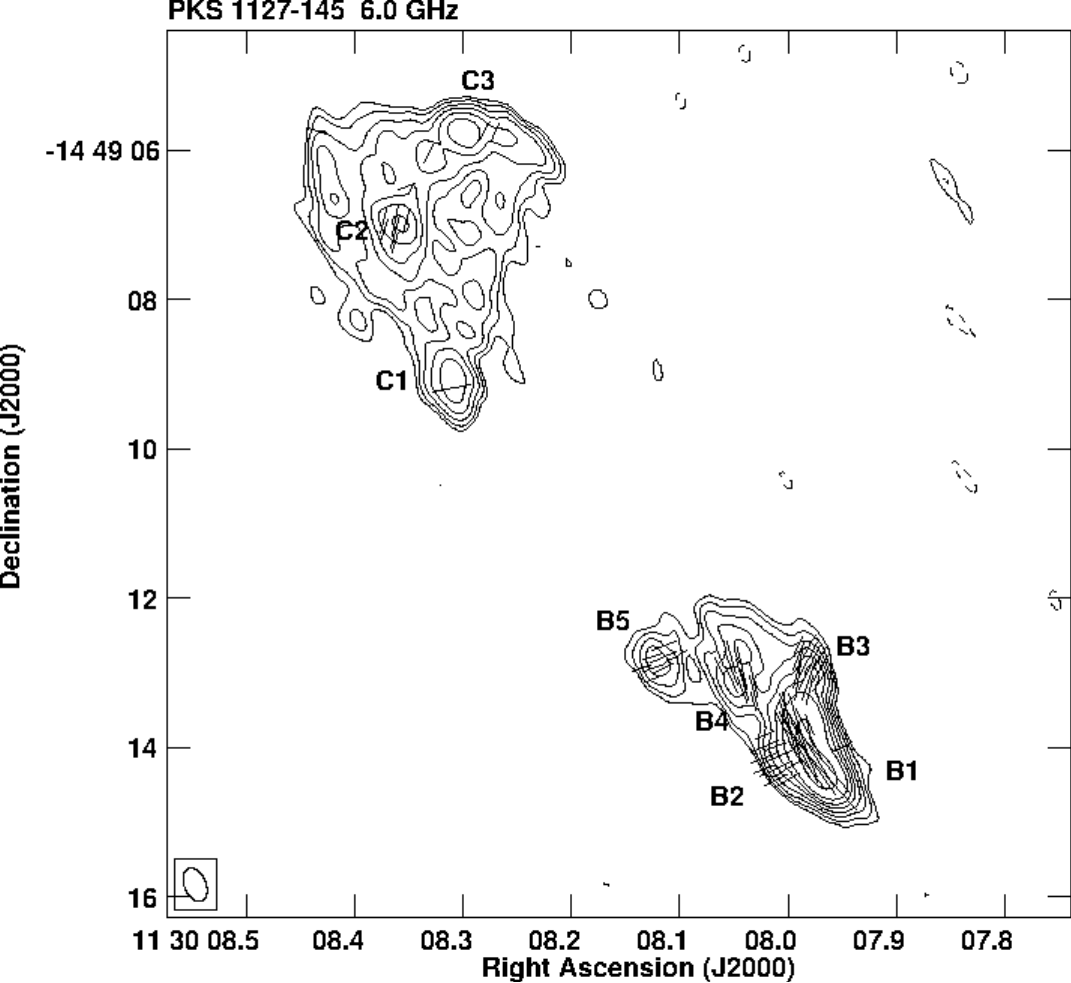}
\includegraphics[width=0.9\columnwidth]{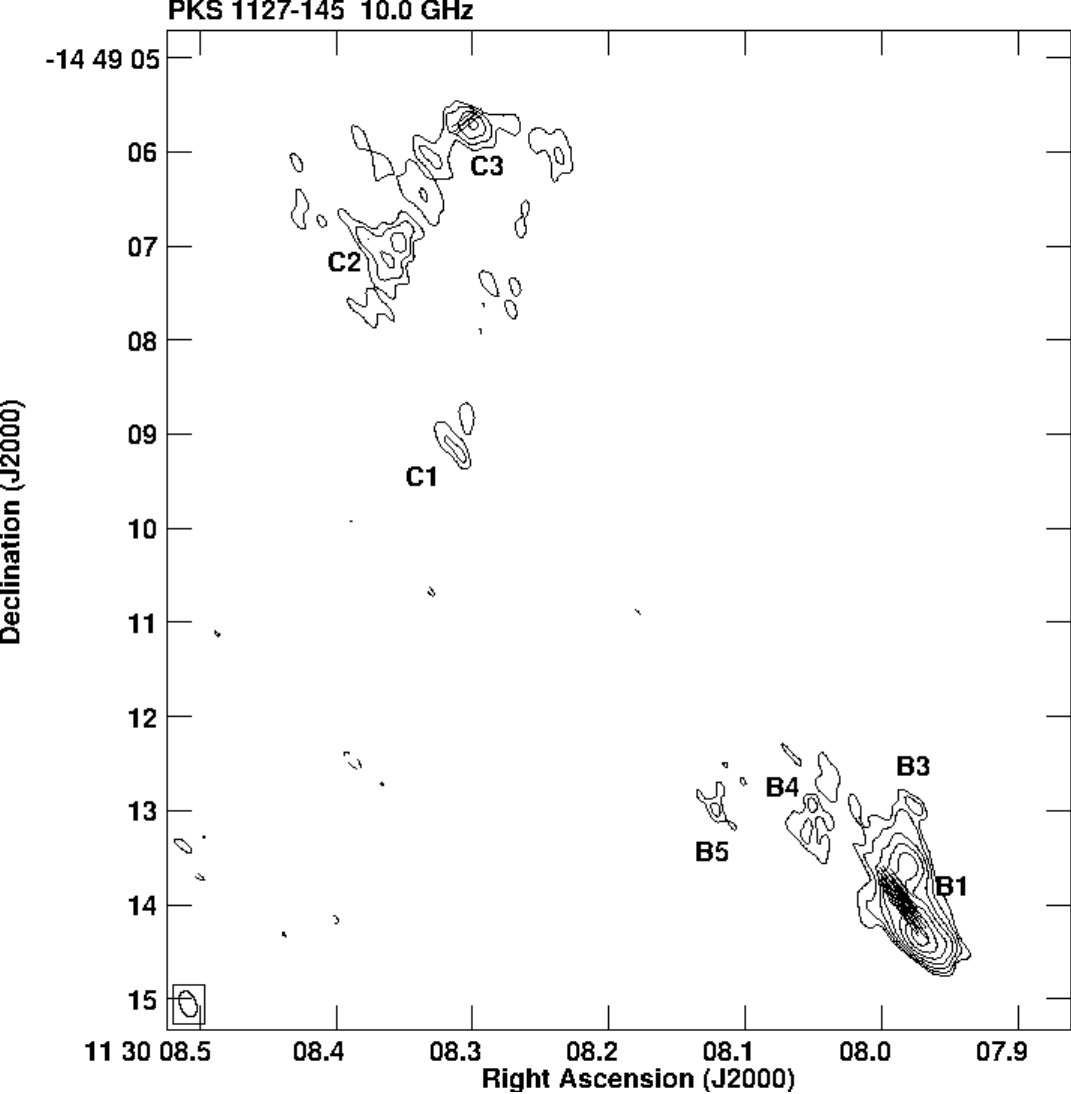}
\caption{VLA total intensity images of the outer regions of PKS\,1127$-$145 at 6 GHz ({\it left}) and 10 GHz ({\it right}). The first contour is 0.05 and 0.045 mJy beam$^{-1}$ at 6
and 10 GHz, respectively, and corresponds to three times the
off-source noise level measured far from the core region. Contours are drawn at [-1,1,1.4,2,2.8,4,5.6,8,16,32,...] times the first contour. Vectors superimposed on the total intensity contours show the position angle of the electric vector, where 0.5-arcsec length corresponds to 0.1 and 0.08 mJy beam$^{-1}$ (polarization intensity) at 6 and 10 GHz, respectively.  } 
    \label{outer-image}
\end{center}
\end{figure*}

\begin{figure*}
\begin{center}
\includegraphics[width=\columnwidth]{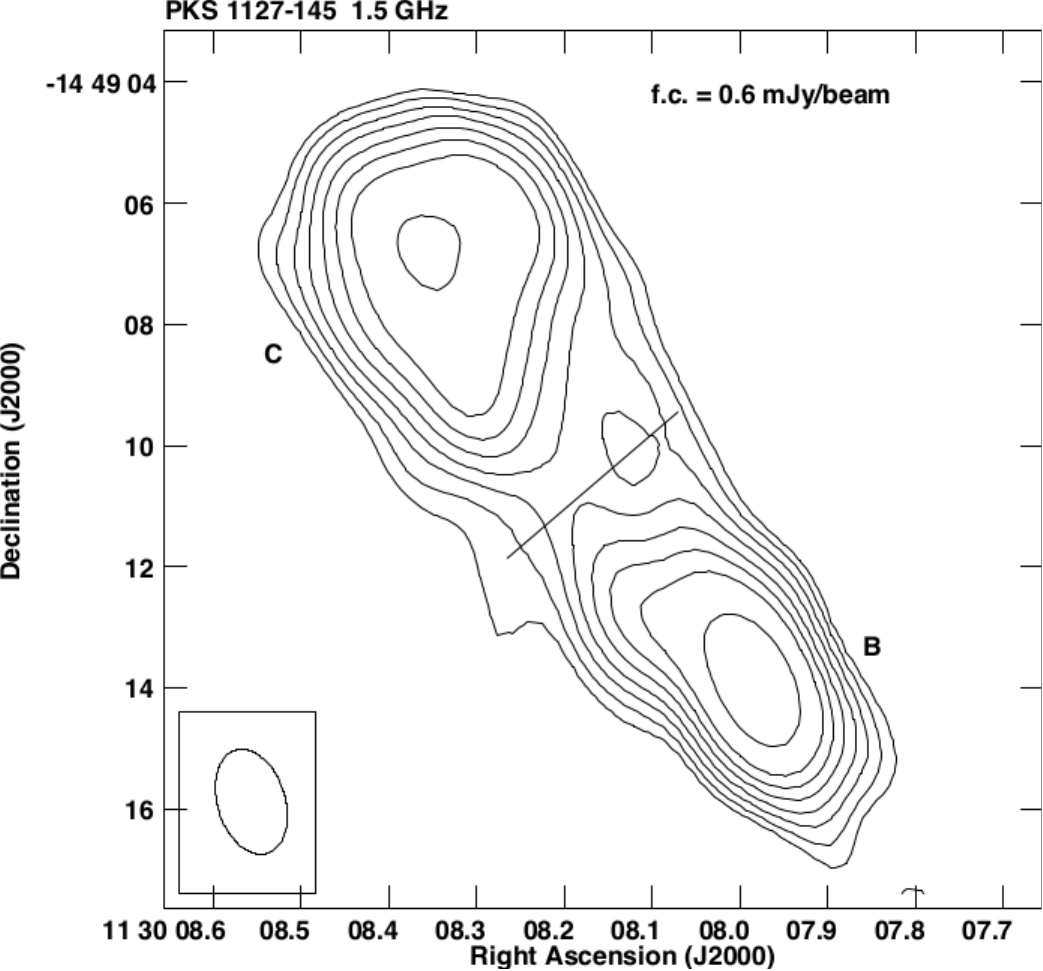}
\includegraphics[width=0.9\columnwidth]{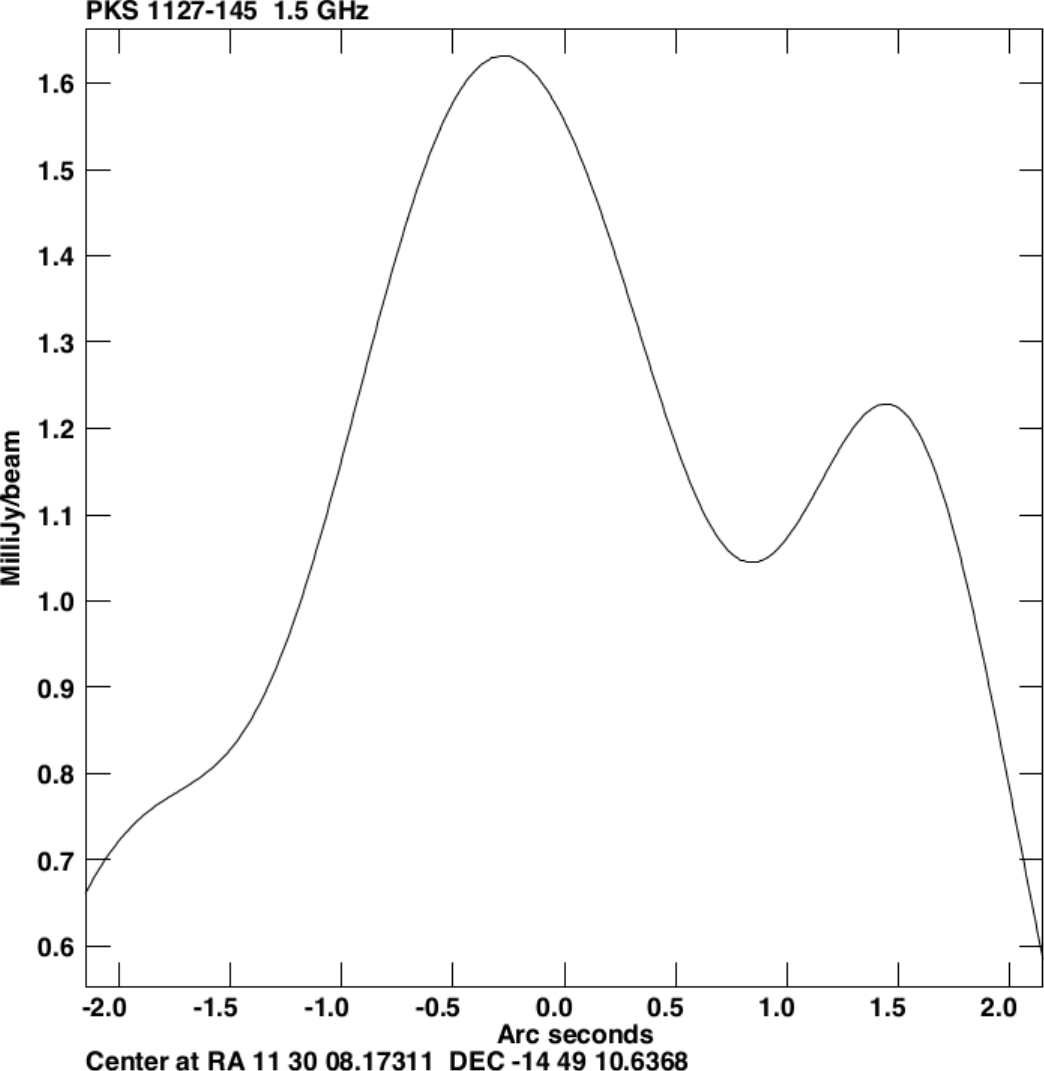}
\caption{VLA total intensity image at 1.5 GHz ({\it left}) of the outer region of PKS\,1127$-145$. The black line indicates the position of the slice used to derive the brightness profile ({right}).}
\label{slice}
\end{center}
\end{figure*}

\begin{table}
\caption{VLA flux density of PKS\,1127$-$145. Column 1: source
  component; columns 2,3,4: flux density at 1.5, 6.0, and 10.0 GHz, respectively.}
\begin{center}
\begin{tabular}{cccc}
 \hline
Comp. & S$_{\rm 1.5}$ & S$_{\rm 6.0}$ & S$_{\rm 10}$\\
     &mJy& mJy & mJy \\
\hline
Core & 5020 $\pm$ 151 & 2492 $\pm$ 75 & 2141 $\pm$ 65 \\ 
I  &    $-$         &  6.0 $\pm$ 0.2 & 4.3 $\pm$ 0.2 \\
O & 1.8 $\pm$ 0.3 &  $-$ & $-$ \\
A    &  5.4 $\pm$ 0.7 &  $-$          &  $-$  \\ 
B    & 37.1 $\pm$ 1.1 & 7.9 $\pm$ 0.3 & 4.7 $\pm$ 0.2 \\
C    & 48.9 $\pm$ 1.5 & 9.1 $\pm$ 0.3 & 4.8 $\pm$ 0.2 \\
 Tot  & 5127 $\pm$ 154 & 2586 $\pm$ 78 & 2156 $\pm$ 65 \\
\hline
\end{tabular}
\end{center}
\label{vla_flux}
\end{table}

\begin{table*}
\caption{Polarization information. Column 1: source component; columns
  2, 4, and 6: polarization flux density in mJy (fractional
  polarization) at 1.5, 6, and 10 GHz, respectively; columns 3, 5, and
  7: EVPA at 1.5, 6.0, and 10 GHz, respectively.}
\begin{center}
\begin{tabular}{ccccccc}
\hline
Comp. & S$_{\rm p, 1.5}$ \;\; ($m_{1.5}$) & $\chi_{\rm 1.5}$ &
S$_{\rm  p, 6}$ \;\; ($m_{6}$) & $\chi_{\rm 6}$ & S$_{\rm p, 10}$ \;\;
($m_{10}$) & $\chi_{\rm 10}$ \\
 & mJy \;\; (\%) & deg & mJy \;\; (\%) & deg & mJy \;\; (\%) & deg \\
(1) & (2) & (3) & (4) & (5) & (6) & (7) \\
\hline
Core & 122 $\pm$ 4 \;\; (2.4) & 46 $\pm$ 3 & 152 $\pm$ 5 \;\; (6.1) &
-26 $\pm$ 3 & 107 $\pm$ 3 \;\; (5.0) & -28 $\pm$ 3 \\
I  &  -  & - & 2.0 $\pm$ 0.1 \;\; (33.3) & -35 $\pm$ 3 & 1.4 $\pm$ 0.1 \;\;
(32.3) & -35 $\pm$ 3 \\
B & - & - & 2.4 $\pm$ 0.1 \;\; (30.4) & 43 $\pm$ 5 & 1.7 $\pm$ 0.1 \;\; (36.2) & 50 $\pm$ 5 \\ 
C & - & - & 2.6 $\pm$ 0.1 \;\; (28.6) & -43 $\pm$ 3 & 1.6 $\pm$ 0.1 \;\;
(33.3) & -47 $\pm$ 3 \\
\hline
\end{tabular}
\end{center}
\label{polla}
\end{table*}

\begin{table}
\caption{\chandra\ Observations.}
\begin{center}
\begin{tabular}{cccc}
 \hline
Date & ObsID & Exposure Time \\ 
& & (ksec) \\ 
\hline
2000-05-28 & 866  & 27.3\\
2005-04-25 & 5708 & 105.5 \\
2021-01-01 & 24911 & 13.6  \\
\hline
\end{tabular}
\end{center}
\label{tab:chandra}
\end{table}

\begin{table}
\caption{X-ray Properties of Jet Knots.}
\begin{center}
\begin{tabular}{cccc}
 \hline
Knot  & $\Gamma ^a $ & $f^b_{0.5-7\; \rm keV}$ \\ 
\hline
I  &   $1.56\pm0.12$ &  $28.3^{+2.8}_{-2.5}$  \\
O  &   $1.40\pm 0.28$ &  $6.3^{+1.4}_{-1.2} $  \\
A  &   $1.76\pm 0.23$  & $7.8^{+1.23}_{-1.0}$  \\ 
B  &   $1.74\pm 0.26$  & $5.9^{+1.3}_{-0.8}$   \\
C  &   $1.49\pm 0.34$  &  $ 4.2^{+1.6}_{-0.9}$ \\ 
\hline
\end{tabular} 
\end{center}
Notes: $^a$ An absorbed power law model with Galactic absorption $N_H=4.09\times 10^{20}$~cm$^{-2}$; $^b$ unabsorbed flux in units of $10^{-15}$ erg~cm$^{-2}$~s$^{-1}$.
\label{tab:knots}
\end{table}

\section{Results}
\label{sec:results}

In this paper we follow the component nomenclature used in
\citet{siemiginowska02} and \citet{siemiginowska07}. The main knots,
A, B, and C are located $\sim 12''$, $18''$ and $27''$ from the core,
respectively. Two additional brightenings, labelled I and O, are
located between the core and knot A (Fig. \ref{vla-image}). 

The flux densities in L band of the main knots are consistent with the values reported in \citet{siemiginowska07}. The lower values in C and X bands are likely related to different characteristics of the observations. The data sets presented in \citet{siemiginowska02} and \citet{siemiginowska07} were obtained with a longer observing time and a more compact VLA configuration that is more effective in picking up diffuse emission than our observations.

\subsection{The 2-arcsec inner jet structure}
\label{sec:innerjet}

The new VLA radio observations allow a detailed analysis of the inner 2-arcsec jet structure\footnote{The pc-scale jet structure imaged by the Very Long Baseline Array at mas-scale resolution will be presented in a forthcoming paper.}, for earlier VLA observations had a resolution not adequate to disentangle it from the core
\citep{siemiginowska07}. 
At 1.5 GHz the inner jet (labelled I in Fig. 1) is slightly resolved and is connected to the outer part of the jet by a low-surface brightness emission that bridges knots O and A.
At 6 and 10 GHz the inner jet is clearly resolved, and extends to about 1.8 arcsec ($\sim$15 kpc) from the core with a position angle of $\sim$70$^{\circ}$ (Fig. \ref{inner-image}), well aligned with the pc-scale jet \citep{jorstad17}. The fractional polarization is about 33 per cent at 6 and 10 GHz, and the EVPA is $\sim -$35$^{\circ}$ at both frequencies indicating no significant Faraday Rotation. Imaging artifacts are clearly visible in total intensity and polarized emission.
No polarized emission is detected at 1.5 GHz outside the core region. This may be due to the
poor S/N reached at this frequency, although some beam
deporalization cannot be ruled out. 

The low fractional polarization of the core (between 2 and 6 per cent, depending on the observing frequency) is typical for the central regions of blazars on arcsecond and milliarcsecond scales \citep[e.g.,][]{odea88,laurent93,lister05,marscher02,harris17,baghel24}.

\subsection{The kpc-scale radio structure}
\label{sec:radio-jet}

The kpc-scale radio structure of
PKS\,1127$-$145 extends for about 30 arcsec ($\sim$250 kpc projected) and is well resolved into several components. 
Super-posed on the faint extended jet emission we observe three knots (labelled O, A, and B in Fig. \ref{vla-image}) before the jet termination at component C.
At about 8 arcsec from the core, the jet slightly bends to a PA $\sim$50$^{\circ}$, in agreement with what is reported in \citet{siemiginowska02}. A further bend to PA$\sim$ 40$^{\circ}$ is observed in correspondence to the outer part of component B marked by component B3 in Fig. \ref{outer-image}.  
Components O and A, at 7.5 and 12 arcsec from the core, respectively, are detected only in L band, suggesting a steep spectral index. We note that component A was detected in the VLA data at 4.9 GHz presented in \citet{siemiginowska07}. As mentioned above, those data were obtained at a lower frequency with a much longer exposure time, and when the array was in a more compact configuration, thus sensitive to emission extending on larger angular scales than those recoverable by our observations. The non-detection of component A suggests that its emission is diffuse on scales larger than $\sim$ 5 arcsec (i.e. $\sim$ 40 kpc).

Moving farther out, at about 18 arcsec from the core there is component B, detected at all frequencies, that marks a re-brightening of the radio emission, while X-rays fade away. Component B has a fractional polarization of about 20\%, similar to the polarization percentage usually found in jet knots \citep[e.g.,][]{bridle94}. The EVPA is $\sim$ 40$^{\circ}$ at the position of component B1 and B4, roughly parallel to the jet axis at the peak, while it is perpendicular to the total intensity contours at the edges of the component (regions B2, B3, and B5).
The EVPA parallel to the jet axis is the opposite of what is found in the inner 2-arcsecond jet region, pinpointing a 90-degree tilt of the magnetic
field. 

In the 1.5-GHz image, component B is connected to component C by diffuse emission, showing a limb-brightened structure as suggested by the brightness profile obtained by interpolating a slice roughly perpendicular to the jet axis with the AIPS task {\tt SLICE} (Fig. \ref{slice}).

The polarized regions B4/B5 and B3 might mark the starting point of the two filaments.
Despite well imaged at 1.5 GHz, at 6 GHz component C is resolved into
several polarized clumps with different EVPA, enshrouded by diffuse
emission. On the other hand, at 10 GHz we could detect only sub-components C2 and C3 and a hint of C1,
while the diffuse emission could be barely imaged, likely due to a combination of sensitivity
and largest recoverable angular scale (Fig. \ref{outer-image}).

\subsection{The kpc-scale X-ray jet}
\label{sec:xray-jet}

The X-ray jet is well aligned with the 30\arcsec\ radio jet 
showing strong X-ray emitting knots, O and A, leading to more prominent radio
knots, B and C. The X-ray knots A and B are connected by the continuous faint X-ray emission. Although faint X-ray emission from knot C is clearly present, no diffuse emission connecting knots B and C is detected 
(see Fig.~\ref{fig:xray-jet}). Overall the X-ray surface brightness declines with the distance from the core. 

We extracted X-ray spectra of the knots for each observation and fit them separately and then simultaneously by applying an absorbed power law with a Galactic absorption column of $N_H = 4.09 \times 10^{20}$~cm$^{-2}$. The best-fit photon index and an unabsorbed 0.5-7\,keV flux resulting from the fit to each knot are presented in Table~\ref{tab:knots}. The results of the simultaneous fitting are consistent between each observations. 
No significant flux increase is detected in the new {\it Chandra} observations. The observed knots' fluxes are consistent between each observations and within the uncertainties listed in Table~\ref{tab:knots}. The available data are not sensitive to small flux variations at the level of $10^{-14}$\,erg~cm$^{-2}$~s$^{-1}$, but we can exclude a flux increase by a factor of 100 in the knots that would dominate over the flux of the core.

\begin{figure*}
    \centering
     \includegraphics[width=\columnwidth]{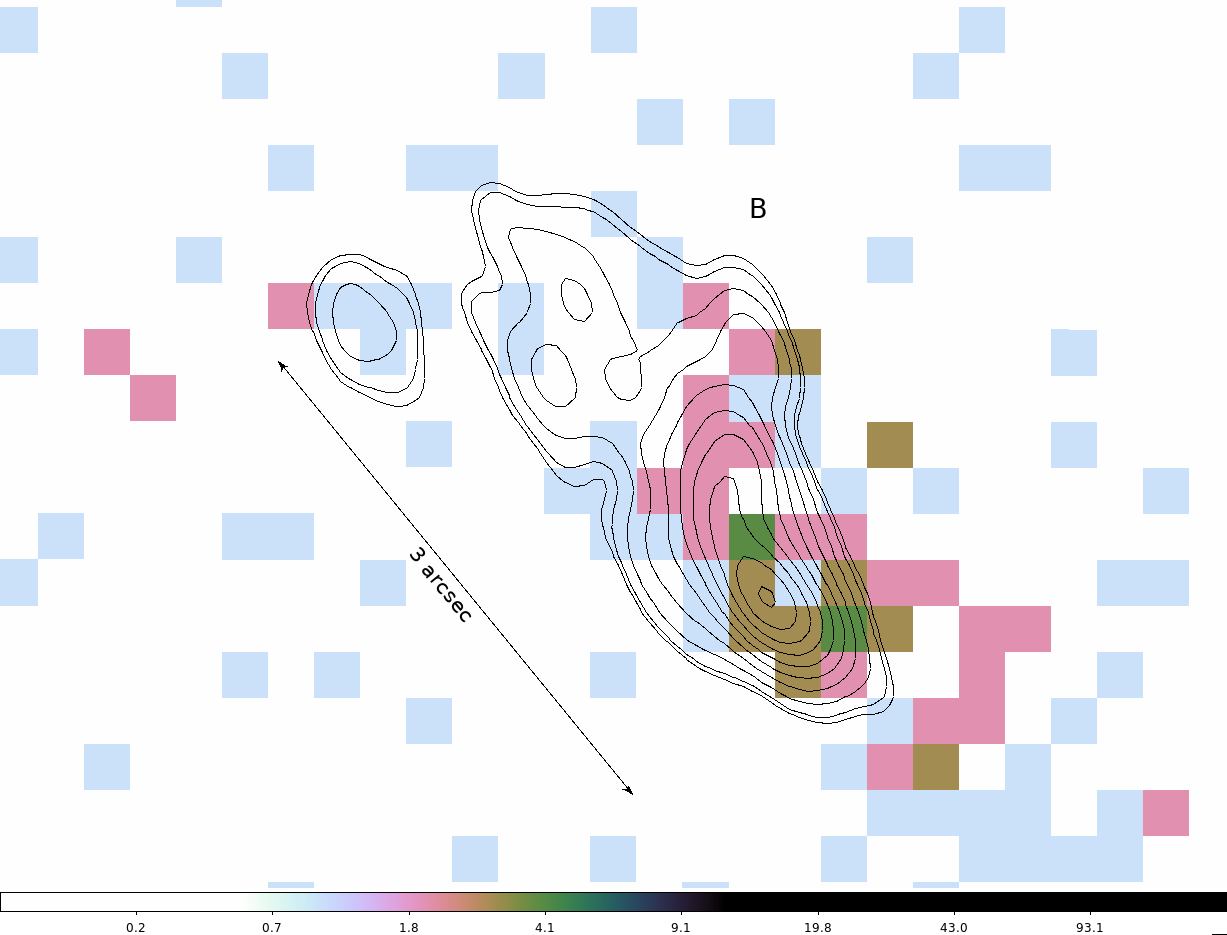}
     \includegraphics[width=\columnwidth]{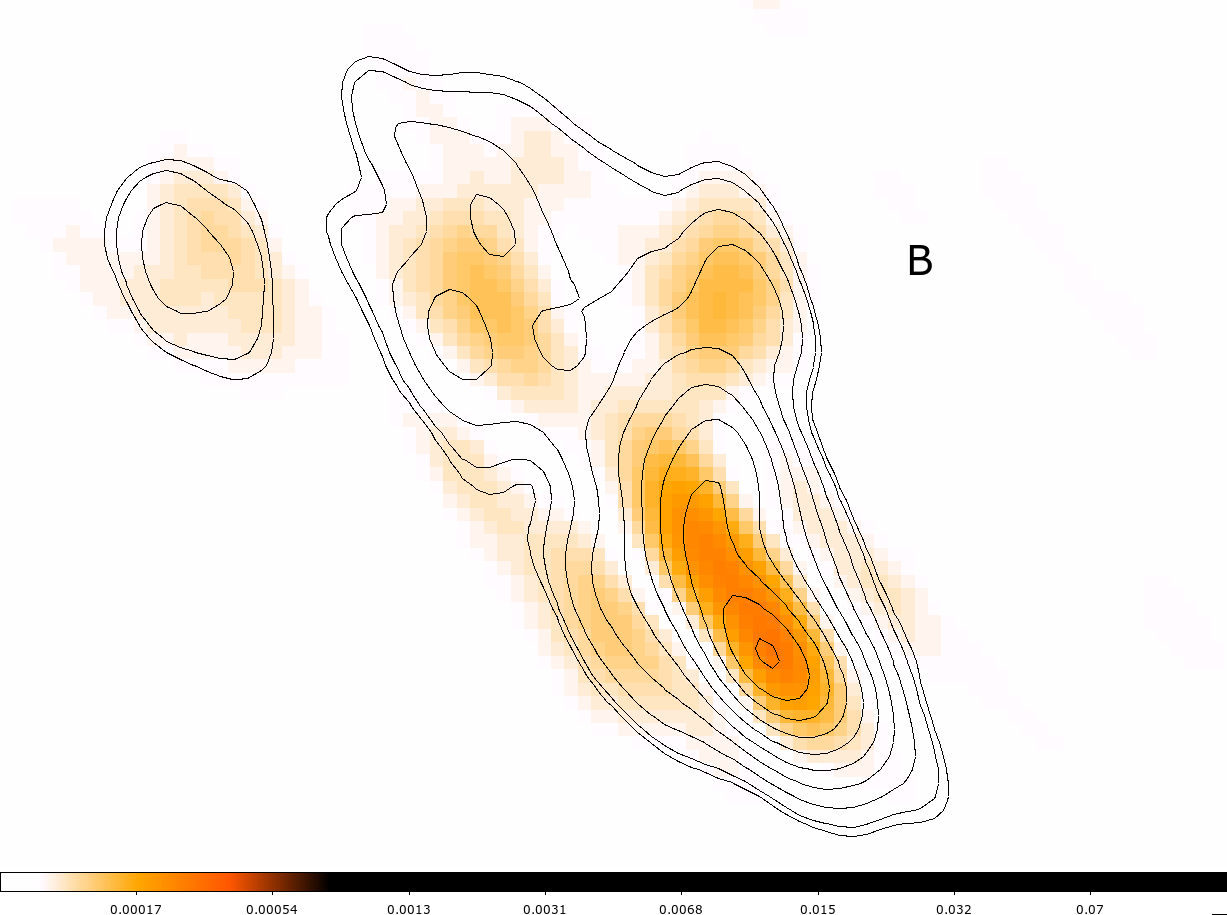}
        \caption{{\it Left}: ACIS-S 0.5--7 keV mosaic image of knot B binned
          to 0.249\arcsec\ pixels. The color map shows the number of
          counts per image pixel. The contours are 6\,GHz VLA data
          starting at 0.075 mJy beam$^{-1}$ with a $\sqrt2$ scale. The
          arrow marks a 3\arcsec\ scale size.
                  {\it Right}: C-band polarization map with the 6\,GHz radio intensity contours as on the left. The color bar shows the polarization with orange marking the strong linear polarization.}
    \label{fig:knotB}
\end{figure*}

\begin{figure*}
    \centering
        \includegraphics[width=\columnwidth]{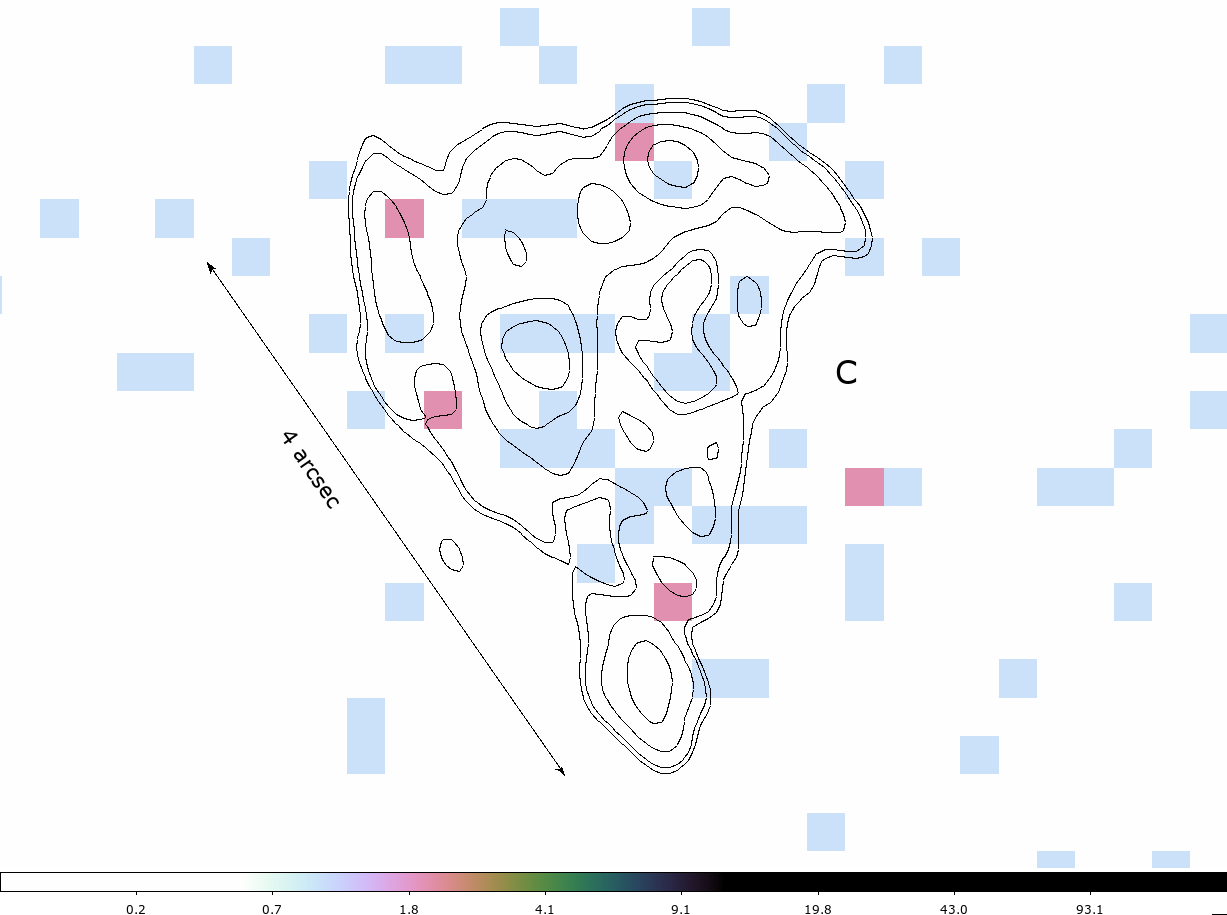}
    \includegraphics[width=\columnwidth]{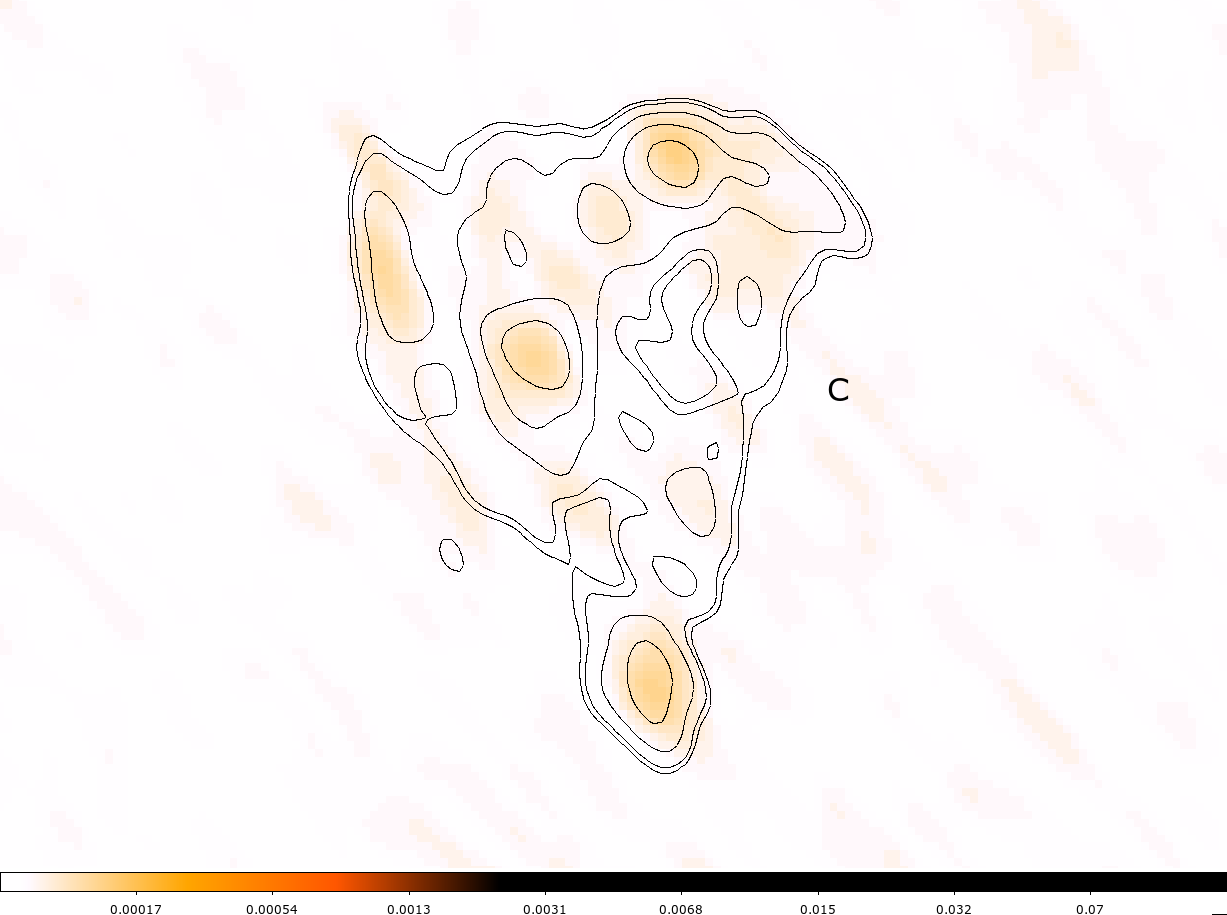}
        \caption{
    Left: ACIS-S  0.5--7 keV mosaic image of knot C binned to 0.246\arcsec\ pixels and 6\,GHz contours starting at 0.075 mJy beam$^{-1}$ with a $\sqrt2$ scale. The color map shows the number of counts per image pixel. The arrow marks a 4\arcsec\ scale size.  
    Right: C-band polarization map with  6\,GHz radio intensity contours starting at 0.075 mJy beam$^{-1}$ ($3\sigma$) with a $\sqrt2$ scale. Compact polarization regions are aligned with the enhanced radio intensity.}
    \label{fig:knotC}
\end{figure*}

Fig.~\ref{fig:knotB} shows the ACIS-S sub-pixel image of knot B. Most of the X-ray counts coincide with the bright and collimated radio part of the knot, but decrease significantly within the fan-like fainter radio emission as the radio jet becomes broader when the X-rays rapidly decline.
A shift between the X-ray and radio emission sites could be observed
and the X-rays seem to precede radio emission, as already claimed in
\citet{siemiginowska02} by comparing X-rays and 1.4-GHz radio
data. Although a misalignment between X-rays and radio emissions seem to be present
  the current image does not provide statistically
significant measurements of a possible shift.
An offset of $\sim$0.5 arcsec was also
claimed by \citet{reddy23}, who made use of the Low-count Image
Reconstruction Algorithm. However, the radio data used by
\citet{reddy23} for investigating the offset have a lower resolution
than ours, and the radio peaks may not coincide when different angular
resolution are considered, owing to the resolved structure of the
component. The same reasoning applies when assessing the offset of
component C.
The X-ray emission of knot C shown in Fig.~\ref{fig:knotC} is spread uniformly across the diffuse radio structure. The linear polarization map (at 6~GHz) shows a few clumps of enhanced polarization which seem to be roughly coincident with the stronger X-ray emission. However, the X-ray S/N is very low for any detailed analysis of the knot B and C regions. 

\subsection{Quasar Core in X-rays}
\label{sec:core}

We extracted the quasar core spectrum from the circular source region ($r=2\arcsec$, corresponding to
95\% PSF) centered on RA=11:30:07.11, Dec=$-$14:49:27.1 
in the most recent \chandra\,observation (ObsId 24911). We fit an absorbed power law model to the spectrum, assuming the Galactic absorption of $N_H = 4.09\times 10^{20} \rm cm^{-2}$, and obtained the best-fit photon index of $\Gamma=1.28\pm 0.05$ and an unabsorbed 0.5--7\,keV flux of  
$(5.98^{+0.22}_{-0.15}) \times10^{-12}$\,ergs~s$^{-1}$~cm$^{-2}$. The corresponding 2--10\,keV flux for the best fit model is ($6.24\pm0.57)\times10^{-12}$\,ergs~s$^{-1}$~cm$^{-2}$.

\section{Discussion}
\label{sec:discuss}

\subsection{The emission mechanism}

Although IC-CMB process often explains the X-ray emission in kpc-scale jets at low and high redshift \citep[e.g.,][]{tavecchio00,sambruna04,erlund06,cheung12,simionescu16,schwartz20,migliori22}, there are some cases where it clearly fails to reproduce the multi-band observations and different scenarios have been proposed \citep[see e.g.,][]{hardcastle06,clautice16,marshall18,tavecchio21,meyer23,reddy23}.

Past studies of radio and X-ray emission from the extended jet of PKS\,1127$-$145 could not unambiguously unveil the dominant mechanism at the origin of its high energy emission \citep{siemiginowska07}.  
The X-ray brightness distribution and the offset between the emission peaks in the two bands challenge the standard one-zone leptonic models. The radio and X-ray spectra pointed to more complex jet radiation processes associated with, for example, a `jet-sheath' structure, or multiple epochs of quasar jet activity \citep{siemiginowska07}. However, the study was limited by the need of averaging over unknown jet sub-structures that could not be resolved by earlier radio observations.

Our full-polarization radio data allow us to investigate for the first time with high-angular resolution both the total intensity and polarized emission along the kpc-scale jet. A decrease of X-ray emission as moving away from the core accompanied by an increase of the radio emission, though not common, has been observed in other kpc-scale X-ray jets, like 3C\,273 \citep{marshall01}, 1136$-$135 and PKS\,1510$-$089 \citep{sambruna04}, and 0827+243 \citep{jorstad04}. In 0827+243 (OJ\,248) the X-ray emission shows an apparent $\sim 90^{\circ}$ bend at about 5 arcsec from the core, and then fades away. On the other hand, at 5 and 15 GHz the large-scale jet is detected only from the bend up to the jet termination. The bend may correspond to a standing shock wave caused by, e.g., a jet-cloud interaction that originates a deflection and a deceleration of the jet flow \citep{jorstad04}. A similar scenario may apply to PKS\,1127$-$145 where component B may represent a standing shock, after which the jet decelerates. The widening of the jet and its possible limb-brightened structure observed beyond component B support a change in the collimation and velocity of the jet, that may explain the different behaviour of the radio and X-ray emission \citep{gk04}. The limb-brightened structure might indicate the co-existence of both longitudinal and transverse velocity gradients at the jet bending.

Although mean EVPA in quasar jets is usually perpendicular to the jet direction up to the jet termination where it becomes parallel, it may also change along the jet in presence of shocks \citep[see e.g.,][]{bridle84, bridle94,marscher02,pushkarev23}. This is what is clearly observed in component B of PKS\,1127$-$145, where the magnetic field becomes perpendicular to the jet axis. A change of EVPA orientation, from perpendicular to parallel to the jet axis, is also observed along the jet of PKS\,1510$-$089 in correspondence of a jet bending \citep{odea88}.
The abrupt rotation of the EVPA may be due to the compression of a tangled magnetic field to a plane perpendicular to the jet axis by a shock \citep{laing80}.
In 0827+243 polarized emission is detected only at the jet termination, precluding the information on the EVPA along the kpc-scale jet.  
Neither a significant change of the EVPA nor a clear bending is observed in the kpc-scale jet of 3C\,273 at the position of X-rays dimming \citep{perley17b}. It is worth mentioning that jet bends are not uncommon in the outer regions of relativistic jets \citep[e.g.,][]{bridle94} and they are detected also in some kpc-scale jets showing the same morphology in radio and in X-rays emissions \citep[e.g. 0723+679 and 1150+497,][]{sambruna04}, though polarization information is in general unavailable. 

A different situation may be represented by component C, where X-ray emission is faint and diluted on a large region. The high angular resolution of our radio data points out a similar situation in the radio band, where the emission is mainly diffuse on several tens of kpc, while only a marginal part comes from (polarized) compact clumps. Our measurements of the radio polarization in the two outer knots are extremely interesting as they highlight strong organized magnetic fields and potential sites of particle acceleration at hundred kpc
distances from the core. This may signify a possibility for the synchrotron X-rays in large scale jets, as it is also suggested for hotspots of radio galaxies \citep[e.g.,][]{hardcastle04,mo20,migliori20}. \\
Multi-frequency full-polarization observations of a sample of kpc-scale X-ray jets are necessary for drawing a more complete view on the physics in these extreme jets.

\subsection{The X-ray flaring region}

Locating the high-energy flaring region in blazars is not trivial. In the shock-in-jet scenario a disturbance may be produced at the jet base, and outbursts may take place at different location, as the disturbance propagates down the jet and encounters (quasi-)stationary features. The manifestation of such moving shocks are superluminal jet knots observed on parsec scale by VLBI observations 
\citep[see e.g.,][]{marscher10,agudo11,casadio15,jorstad17,lister18,mo20b,lico22,kramarenko22,weaver22}. 
The discovery in the nearby radio galaxy M87 of a high-energy flare taking place in the jet component HST-1 clearly proves that outbursts as bright as the core itself can take place hundred of parsecs away from the central engine \citep{cheung07,harris2009}. An X-ray flux increase was observed in the jet of Pictor\,A at about 35 kpc from the core \citep{marshall10,hardcastle16a}. However, in this case the enhancement did not achieve the luminosity of the core of Pictor\,A.

Neither in X-rays nor at radio frequencies we observe any brightening
of a jet knot in PKS\,1127$-$145. Despite short, {\it Chandra}
observations would have been deep enough to detect the flare if it had
been from the kpc-scale X-ray jet. On the other hand, the X-ray flux
of the quasar core is roughly 2.6 times higher than that reported in
\citet{siemiginowska02} and comparable to that observed by {\it Swift}
just after the flaring episode \citep{dammando20b}, locating the
flaring region within the VLA core.
 We notice that at the redshift of PKS\,1127$-$145 the X-ray flaring component observed in the jet of Pictor\,A would be at about 4 arcsec from the core (i.e. closer than component O), while M87 HST-1 would be at 12 milliarcsecond from the core, in the innermost pc-scale structure.
 Multi-epoch very long baseline array observations of the central parsec-scale jet region will be presented in a dedicated paper focusing on the multi-band analysis of the source core.

\section{Conclusions}
\label{sec:con}

We presented results on {\it Chandra} X-ray observations and subarcsecond polarimetric VLA observations of PKS\,1127$-$145 performed during a flaring event detected in $\gamma$-rays by {\it Fermi}-LAT. The conclusions we can draw from this study are:

\begin{itemize}

\item The high angular resolution of the new VLA data allowed us to image the inner kpc jet for the first time. The inner jet is highly polarized and the magnetic field is parallel to the jet axis;\\

\item In agreement with earlier observations, the outer knots are the brighter in radio, contrary to what is found in X-rays. A re-brightening of the radio emission is observed at about 150 kpc (projected) from the core where the jet slightly bends and likely decelerates. The magnetic field at the position of the bend shows a 90-degree rotation, likely due to compression to a plane perpendicular to the jet axis, as it is observed in many knots of relativistic jets, though usually on pc-scale;\\

\item The outermost component is resolved into several polarized compact regions enshrouded by diffuse emission. Such patchy structure, reminiscent of some hotspots in radio galaxies, should be kept into consideration when modelling the spectral energy distribution of these kpc-scale structures;\\

\item The limb-brighetened structure and the widening of the jet connecting component B to the jet termination support a deceleration and decollimation of the jet flow in the outer part of PKS\,1127$-$145;\\

\item The X-ray flux from the quasar core is consistent with 
the {\it Swift} measurements during the 2020-2021 flaring period, i.e. 2.6 times higher than the flux in the 2000 data reported by \citet{siemiginowska02}. 
Neither radio nor strong X-ray flux variability
is observed from any region of the kpc-scale jet at the level detected by 
{\it Swift}.
This strongly indicates that the high-energy flaring episode was located in the VLA core.\\

\end{itemize}

The faint X-ray emission from the outermost knots prevents a detailed study of the correlation between radio and X-ray emission at the edge of the jet. Deep X-rays observations are then necessary to investigate whether the X-ray morphology is consistent with the radio structure, with clumps and filaments, and to determine any alignment and/or separation of individual features in the bands. Recently, the high angular resolution of Low Frequency Array observations in the MHz regime, pointed out that the flux density at 150 MHz of the knots in the jet of 4C\,+19.44 are below the values expected by extrapolating the GHz spectra, suggesting a low-energy curvature of the particle energy distribution \citep{harris19}. The jump in resolution and sensitivity in the MHz regime that will be achieved with the advent of the square kilometre array, together with multi-band information, will provide important information on the electron energy distribution, particle acceleration, and energy dissipation at the periphery of relativistic jets. 
However, the required high angular resolution X-ray imaging  is currently only achievable in the {\it Chandra} observations and it will not be possible until the new generation telescopes, such as {\it Lynx} \citep{Lynx2019} planned mission, are built.

\section*{Acknowledgment}

We thank the anonymous referee for reading the manuscript carefully and making valuable suggestions.
We wish to thank Patrick Slane, Director of the Chandra X-ray Center,
for approving our Director's Discretionary Time (DDT) request, and the
{\it Chandra} team for carrying out the new observations (obsid 24911).
The VLA is operated by the US 
National Radio Astronomy Observatory which is a facility of the National
Science Foundation operated under cooperative agreement by Associated
Universities, Inc. This work has made use of the NASA/IPAC Extragalactic Database (NED) which is operated by the JPL, California Institute of Technology, under contract with the National Aeronautics and Space Administration. 
This research has made use of data obtained from the Chandra Data Archive  and software provided by the Chandra X-ray Center (CXC) in the application packages CIAO and Sherpa. 
A.S. was supported by NASA contract NAS8-03060 (Chandra X-ray Center).

\end{document}